\documentclass[10pt,conference]{IEEEtran}
\usepackage[utf8]{inputenc}
\usepackage[T1]{fontenc}
\usepackage[numbers,sort&compress]{natbib}
\usepackage{booktabs}
\usepackage{balance}
\usepackage{graphicx}
\usepackage{xcolor}
\usepackage{amssymb}
\usepackage{url}
\usepackage{tabularx}
\usepackage{tikz}
\usetikzlibrary{arrows.meta,calc,fit,backgrounds,positioning}
\usepackage[hidelinks]{hyperref}
\usepackage{cleveref}
\usepackage{xspace}
\newcommand{\AgentTorrent}{\emph{AgentTorrent}\xspace}
\newcommand{\AgentConfigs}{\emph{AgentConfigs}\xspace}
\crefname{section}{Section}{Sections}
\Crefname{section}{Section}{Sections}
\crefname{table}{Table}{Tables}
\crefname{figure}{Figure}{Figures}
\crefname{figure}{Figure}{Figures}

\DeclareUrlCommand\fn{\urlstyle{tt}}
\newcommand{\prevConfigured}{4,738}     
\newcommand{\prevMechRows}{15,591}      
\newcommand{\prevCommits}{148,519}      
\newcommand{\absConfigured}{10,944}     
\newcommand{\absMechRows}{62,374}       
\newcommand{\artifactCountMultiple}{four} 
\newcommand{\absPlanningDocs}{27,593}   
\newcommand{\absCommits}{more than 1.2 million} 
\newcommand{\curEndDate}{\mbox{2026-10-22}}    
\newcommand{\curFrame}{46,503}          
\newcommand{\curTrue}{40,901}           
\newcommand{\curFalse}{4,222}           
\newcommand{\curUnclear}{25}            
\newcommand{\curExcluded}{1,355}        
\newcommand{\curCollectionFirst}{\mbox{2026-11-04}} 
\newcommand{\curCollectionLast}{\mbox{2026-11-06}}  
\newcommand{\curConfigured}{10,944}     
\newcommand{\curRowsContext}{25,353}    
\newcommand{\curRowsSkills}{23,585}     
\newcommand{\curRowsRules}{3,483}       
\newcommand{\curRowsSettings}{3,532}    
\newcommand{\curRowsSubagents}{2,713}   
\newcommand{\curRowsCommands}{1,436}    
\newcommand{\curRowsMcp}{885}           
\newcommand{\curRowsPlugins}{751}       
\newcommand{\curRowsHooks}{619}         
\newcommand{\curRowsLsp}{17}            
\newcommand{\curFlagContext}{10,026}    
\newcommand{\curFlagSkills}{3,218}      
\newcommand{\curFlagRules}{680}         
\newcommand{\curFlagSettings}{2,246}    
\newcommand{\curFlagSubagents}{651}     
\newcommand{\curFlagCommands}{281}      
\newcommand{\curFlagMcp}{584}           
\newcommand{\curFlagPlugins}{291}       
\newcommand{\curFlagHooks}{498}         
\newcommand{\curFlagLsp}{16}            
\newcommand{\curNoSpecific}{2,820}      
\newcommand{\curCommits}{1,281,628}     
\newcommand{\curCommitRepos}{9,254}     
\newcommand{\curOtherCommits}{72,527}   
\newcommand{\curOtherRepos}{2,323}      
\newcommand{\curCopilotCommits}{184,386} 
\newcommand{\curCopilotReviewerOnly}{39,047} 
\newcommand{\curGeminiCommits}{6,360}   
\newcommand{\curGeminiReviewerOnly}{5,261} 
\newcommand{\curImports}{10,662}        
\newcommand{\curImportRepos}{2,170}     
\newcommand{\curImportsDeep}{1,551}     
\newcommand{\curImportsDeepRepos}{176}  
\newcommand{\curPlanning}{27,593}       
\newcommand{\curPlanningRepos}{417}     
\newcommand{\curMultiMechRepos}{4,124}   
\newcommand{\curMultiMechPct}{37.7\%}    
\newcommand{\curConfigPlanCommitRepos}{388} 
\newcommand{\curConfigPlanCommitPct}{93.0\%} 
\newcommand{\curPlanningCredited}{36.1\%} 
\newcommand{\curContextCredited}{23.8\%} 
\newcommand{\curAgenticRepos}{123}      
\newcommand{\curSampleOnlySix}{312,453} 
\newcommand{\ghsMinStars}{10}           
\newcommand{\parContributorsMin}{two}   
\newcommand{\parPullsMin}{one}          
\newcommand{\parCreatedMonths}{18}      
\newcommand{\parCommittedMonths}{six}   
\newcommand{\parTopLanguages}{ten}      
\newcommand{\parMedianCommits}{358}     
\newcommand{\parMedianWatchers}{5}     
\newcommand{\parLinkedPages}{five}      
\newcommand{\parMargin}{3\%}            
\newcommand{\parConfidence}{95\%}       
\newcommand{\heuPathRows}{358}          
\newcommand{\heuCommitRows}{431}        
\newcommand{\heuOtherRows}{402}         
\newcommand{\heuCommitDocRows}{17}      
\newcommand{\heuPlanningRows}{523}      
\newcommand{\heuPlanningFrameworks}{73} 
\newcommand{\heuAllRows}{2,343}         
\newcommand{\impCasesClaude}{74}        
\newcommand{\impCasesGemini}{66}        
\newcommand{\watchPages}{438}           
\newcommand{\revCommitsAdded}{22,371}   
\newcommand{\revCommitsRemoved}{275}    
\newcommand{\revNewTools}{87}           
\newcommand{\evSample}{1,039}           
\newcommand{\evAgreed}{1,018}           
\newcommand{\evAgreement}{98.0\%}       
\newcommand{\evAgreementLow}{96.9\%}    
\newcommand{\evAgreementHigh}{98.7\%}   
\newcommand{\evKappa}{0.87}             
\newcommand{\evDisagreements}{21}       
\newcommand{\evRightDeepseek}{11}       
\newcommand{\evRightGlm}{8}            
\newcommand{\evRightNeither}{2}       
\newcommand{\evDeepseekWrong}{10}       
\newcommand{\daPathPrec}{99.87\%}        
\newcommand{\daPathPrecLow}{94.4\%}      
\newcommand{\daPathPrecHigh}{100\%}      
\newcommand{\daPathItems}{806}           
\newcommand{\daCommitPrec}{99.96\%}      
\newcommand{\daCommitPrecLow}{96.6\%}    
\newcommand{\daCommitPrecHigh}{100\%}    
\newcommand{\daCommitItems}{360}         
\newcommand{\daConfSampled}{100}         
\newcommand{\daConfPathsFound}{383}      
\newcommand{\daConfPaths}{386}           
\newcommand{\daUnconfSampled}{200}       
\newcommand{\daUnconfMissed}{one}        
\newcommand{\daMissedHigh}{833}          
\newcommand{\daUnconf}{29,955}           
\newcommand{\daRecallCommits}{1,000}     
\newcommand{\daRecallCredited}{19}       
\newcommand{\daRecallFound}{18}          
\newcommand{\daAttrCorrect}{571}         
\newcommand{\daAttrLabeled}{574}         
\newcommand{\daConfidence}{95\%}         
\newcommand{\daImportPrec}{99.98\%}       
\newcommand{\daImportPrecLow}{90.5\%}     
\newcommand{\daImportItems}{157}         
\newcommand{\daPlanPrec}{99.8\%}          
\newcommand{\daPlanPrecLow}{93.3\%}       
\newcommand{\daPlanItems}{366}           
\newcommand{\daSecondPassShare}{20\%}    
\newcommand{\daPassItems}{650}           
\newcommand{\daPassDisagree}{7}          
\begin{document}
\bstctlcite{IEEEexample:BSTcontrol}

\title{AgentTorrent: A Dataset of Context and Harness Engineering Artifacts on GitHub}

\author{%
\IEEEauthorblockN{Sebastian Baltes\IEEEauthorrefmark{1},
Seyedmoein Mohsenimofidi\IEEEauthorrefmark{1},
Levi B\"{o}hme\IEEEauthorrefmark{2},
Vahram Nersesyan\IEEEauthorrefmark{1},\\
Muhammad Auwal Abubakar\IEEEauthorrefmark{3},
Jai Lal Lulla\IEEEauthorrefmark{4},
Matthias Galster\IEEEauthorrefmark{3},
Christoph Treude\IEEEauthorrefmark{4}}
\IEEEauthorblockA{\IEEEauthorrefmark{1}Heidelberg University, Heidelberg, Germany\quad
\IEEEauthorrefmark{2}University of Bayreuth, Bayreuth, Germany\\
\IEEEauthorrefmark{3}University of Bamberg, Bamberg, Germany\quad
\IEEEauthorrefmark{4}Singapore Management University, Singapore, Singapore\\
{\footnotesize sebastian.baltes@uni-heidelberg.de, s.mohsenimofidi@uni-heidelberg.de, levi.boehme@uni-bayreuth.de, vahram.nersesyan@stud.uni-heidelberg.de,}\\
{\footnotesize muhammad.abubakar@uni-bamberg.de, jailal.l.2025@phdcs.smu.edu.sg, mgalster@ieee.org, ctreude@smu.edu.sg}}}

\maketitle

\begin{abstract}
Over the past months, developers have moved from formulating detailed prompts to engineering the context that agentic coding tools read and to configuring the underlying harnesses. For example, developers commit context files, skills, hooks, and MCP configurations to their repositories, and these files control which instructions a harness reads and which actions it can take. Developers may also commit the specs, plans, and task files that agentic coding tools and spec-driven development frameworks write for agents to implement. The \AgentTorrent dataset records these files for six proprietary and open-source agentic coding tools in \absConfigured{} engineered software projects on GitHub. It contains \absMechRows{} configuration artifacts, the files that these artifacts import, \absPlanningDocs{} agentic planning documents, and \absCommits{} commits that credit an AI tool. All heuristics used for data collection are grounded in tool documentation or derived from GitHub repositories, and we evaluate their detection accuracy with each collection. We use open-weight models to classify whether GitHub repositories host engineered software projects. Unlike most datasets, \AgentTorrent links configuration artifacts, imported files, agentic planning documents, and commits that credit an AI tool within the same repositories. Configuration artifacts, planning documents, and credited commits occur together in \curConfigPlanCommitRepos{} repositories. For example, our dataset allows researchers to study whether a tool is credited in commits that add instructions for that same tool and how instruction files import and contradict each other. Our data collection pipeline supports regular updates and automatically checks vendor documentation and reports unknown artifacts and agentic coding tools. The dataset and the pipeline are open-source and are archived on Zenodo.
\end{abstract}


\section{Introduction}
\label{sec:introduction}

Agentic coding tools such as \emph{Claude Code}, \emph{Codex}, and \emph{Cursor} can implement software changes with a high degree of autonomy~\cite{robbes2026agentic, galster2026configuring}. To guide their work, developers engineer the context that an agentic coding tool reads and configure the harness that governs its actions~\cite{lulla2026loop}. Developers frequently commit artifacts to their repositories that instruct and control agent harnesses through different \emph{configuration mechanisms}~\cite{galster2026configuring}. Such mechanisms include context files, skills, and hooks. We refer to each instance of a configuration mechanism, such as a \fn{CLAUDE.md} file or a skill folder, as a \emph{configuration artifact}.

The instructions that an agentic coding tool reads can extend beyond individual configuration artifacts. A context file can import other files or refer to an \emph{agentic planning document}, such as a specification, plan, or task list that spec-driven development frameworks produce before implementation. Studies of context files have examined how projects instruct agentic coding tools~\cite{mohsenimofidi2026context, chatlatanagulchai2026agent}, but related datasets focus on individual artifacts and use different repository samples and collection dates (\cref{tab:similar}). Therefore, researchers cannot simply merge the datasets to study the adoption of agentic software engineering across repositories and artifact types.

\AgentTorrent brings this evidence together for \curConfigured{} engineered software projects on GitHub. The dataset covers \emph{Claude Code}, \emph{GitHub Copilot}, \emph{OpenAI Codex}, \emph{Cursor}, \emph{Gemini}, and \emph{OpenCode}. It contains configuration artifacts across ten configuration mechanisms, their imported files, agentic planning documents, and commits that credit an agentic coding tool. For example, \AgentTorrent enables researchers to study whether planned tasks are consistent with repository instructions, permissions, and checks, and whether agentic coding tools co-author the instructions and controls for subsequent work.

\AgentTorrent succeeds our previous \AgentConfigs dataset~\cite{galster2026dataset, baltes2026dataset} published in March 2026, for which we collected \prevMechRows{} configuration artifacts for five tools in \prevConfigured{} repositories. The new dataset contains \absMechRows{} configuration artifacts, an increase by a factor of approximately \artifactCountMultiple{}. We added \emph{OpenCode} because our data suggested that it was the most popular tool on GitHub that we had not yet targeted. Our new dataset also adds plugins and language servers as configuration mechanisms, resolves file imports, adds agentic planning documents, and includes commits that credit AI tools other than those we target. Each detection heuristic now cites supporting vendor documentation or examples from GitHub repositories, and our repository classification now uses open-weight models and is complemented by human validation.

\section{Related Datasets}
\label{sec:similar}

\AgentTorrent links configuration artifacts, imported files, agentic planning documents, and credited commits for the same repositories, supporting studies of how tools contribute to the instructions and plans for subsequent work.

\begin{table}[t]
\centering
\footnotesize
\caption{Related datasets and their repository selection criteria.}
\label{tab:similar}
\setlength{\tabcolsep}{3pt}
\begin{tabularx}{\columnwidth}{@{}>{\raggedright\arraybackslash}p{2.2cm}>{\raggedright\arraybackslash}X>{\raggedright\arraybackslash}X@{}}
\toprule
Dataset & Content & Selection of repositories \\
\midrule
\emph{GitSkills}~\cite{destefanis2026gitskills} & 3,797,117 \fn{SKILL.md} files & code search by file name \\
\emph{SpecMine}~\cite{agarwal2026specmine} & 470,795 specs files, 98,574 \emph{Kiro} files & code search by file names \\
\emph{GHAgentFiles}~\cite{cardoen2026ghagentfiles} & 142K+ file histories: context files, skills, subagents & \emph{GHS}, 1+ year old, not a fork, 300+ commits, active in the last year \\
\citeauthor{zarzycki2026census}~\cite{zarzycki2026census} & root \fn{AGENTS.md} or \fn{CLAUDE.md} in 6,006 of 22,854 repositories & 1,000+ stars, pushed in the last 90 days, not a fork or archived, 13 languages \\
\emph{MCP-Eco}~\cite{tounoussi2026mcpeco} & 9,200 MCP declarations, 78,302 candidate servers & code search for MCP servers, dir., registries \\
\citeauthor{robbes2026agentic}~\cite{robbes2026agentic} & configuration files, authors, branches, or labels: 28,418 of 128,018 projects & \emph{GHS}, no forks, 5,000+ LOC, 100+ commits, active in the last 3 months \\
\citeauthor{khosravani2026census}~\cite{khosravani2026census} & configuration files, commit messages, authors, bots: 180M+ repositories & \emph{World of Code} without fork copies \\
\emph{AIDev}~\cite{li2026aidev} & 932,791 pull requests; five agentic coding tools & GitHub search query per agentic coding tool \\
\AgentConfigs~\cite{galster2026dataset} & \prevMechRows{} configuration artifacts, \prevCommits{} commits & \emph{GHS}, filters, engineered software projects \\
\midrule
\AgentTorrent & \absMechRows{} configuration artifacts, \curImports{} imports, \curPlanning{} agentic planning documents, 1.2M+ credited commits & \emph{GHS}, \ghsMinStars{}+ stars, OSI license, \parTopLanguages{} languages; forks/archives excluded; contribution/activity filters; eng. software projects \\
\bottomrule
\end{tabularx}
\end{table}

\cref{tab:similar} lists related datasets and how repositories were selected. The authors of \emph{SpecMine}, for example, acknowledge that searching by file name mixes ``\emph{serious projects with tutorials, template clones, and one-off experiments}''. Our pipeline supplements activity thresholds with a model that checks project purpose and engineering practices (\cref{sec:sampling}). Selecting repositories independently of configuration also provides a denominator for prevalence: \curConfigured{} of the \curTrue{} repositories in our sample contain a detected configuration artifact. Among the datasets, \AgentTorrent is the only one that records configuration artifacts, imports, agentic planning documents, and credited commits for the same repositories.

Interpreting artifact frequency and commit counts requires checking both whether detected records are correct (precision) and whether relevant records were missed (recall). \citeauthor{robbes2026agentic}, \citeauthor{khosravani2026census}, and the authors of \emph{GHAgentFiles} estimated detection precision from labeled samples; \citeauthor{khosravani2026census} state that ``\emph{absolute recall is unknown}''. We evaluated both precision and recall, using repository file lists and complete commit logs to search beyond detected files and commits (\cref{sec:validation}).

\section{Data Description and Research Use}
\label{sec:overview}
\label{sec:use-cases}

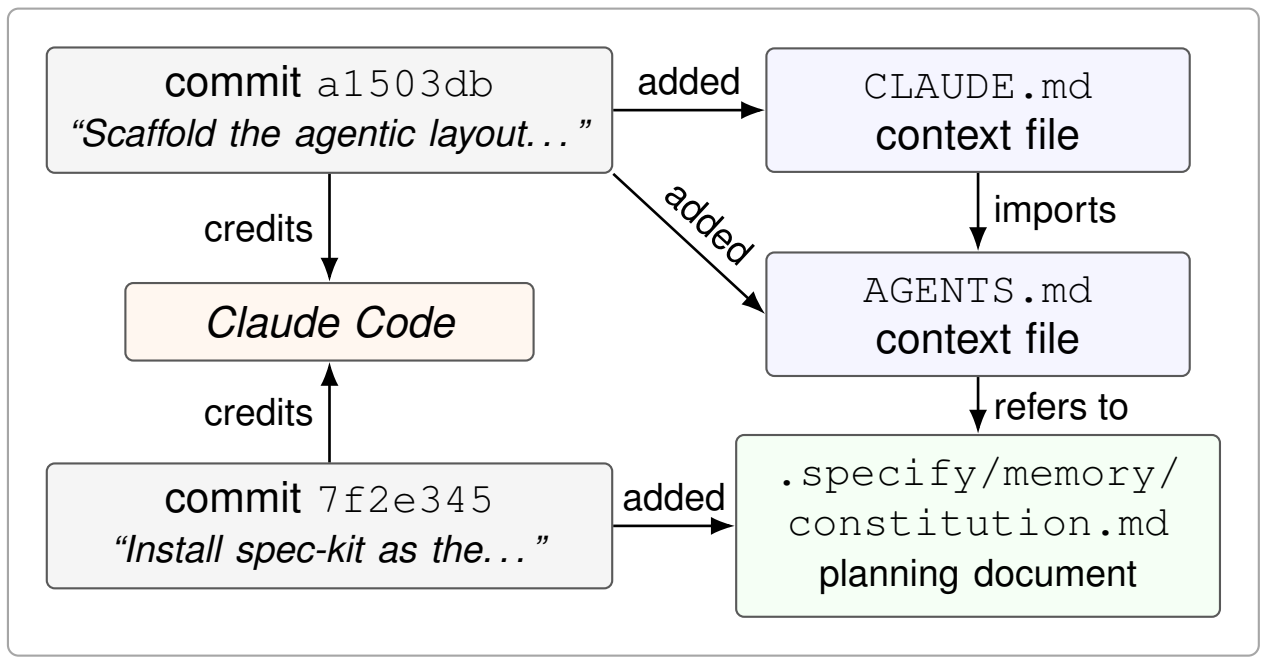
\begin{figure}[t]
\centering
\begin{tikzpicture}[
  font=\sffamily\fontsize{8}{9.5}\selectfont,
  file/.style={draw=black!65,rounded corners=1.5pt,fill=blue!4,align=center,inner sep=1.5mm,text width=2.5cm},
  commit/.style={draw=black!65,rounded corners=1.5pt,fill=black!4,align=center,inner sep=1.2mm,text width=3.5cm},
  product/.style={draw=black!65,rounded corners=1.5pt,fill=orange!6,align=center,inner sep=1.5mm,text width=2.4cm},
  link/.style={-{Latex[length=1.6mm]},line width=0.5pt},
  relation/.style={font=\sffamily\fontsize{7}{8}\selectfont,inner sep=0.7pt}
]
\node[commit] (instructions) at (0,0) {commit \texttt{a1503db}\\{\fontsize{7}{8}\selectfont\mbox{\emph{``Scaffold the agentic layout\ldots''}}}};
\node[file] (claude) at (4.29,0) {\texttt{CLAUDE.md}\\context file};
\node[file] (agents) at (4.29,-1.35) {\texttt{AGENTS.md}\\context file};
\node[product] (claudecode) at (0,-1.4) {\emph{Claude Code}};
\node[commit] (planning) at (0,-2.75) {commit \texttt{7f2e345}\\{\fontsize{7}{8}\selectfont\emph{``Install spec-kit as the\ldots''}}};
\node[file,fill=green!4,text width=2.9cm] (plan) at (4.29,-2.75) {\texttt{.specify/memory/}\\\texttt{constitution.md}\\{\fontsize{7}{8}\selectfont planning document}};
\draw[link] (instructions) -- node[relation,above=2pt] {added} (claude);
\draw[link] (instructions.south east) -- node[relation,sloped,above=2pt] {added} (agents.west);
\draw[link] (planning) -- node[relation,above=2pt] {added} (plan);
\draw[link] (instructions) -- node[relation,left=2pt] {credits} (claudecode);
\draw[link] (planning) -- node[relation,left=2pt] {credits} (claudecode);
\draw[link] (claude) -- node[relation,right=2pt] {imports} (agents);
\draw[link] (agents) -- node[relation,right=2pt] {refers to} (plan);
\begin{scope}[on background layer]
\node[draw=black!35,rounded corners=2pt,fit=(claude)(agents)(instructions)(plan)(planning)(claudecode),inner sep=2.5mm] {};
\end{scope}
\end{tikzpicture}
\caption{Motivating example: Linked files and commits in \href{http://github.com/nornir-automation/nornir}{\texttt{nornir-automation/nornir}}. The reference to the planning document appears in the collected \texttt{AGENTS.md} file.}
\label{fig:links}
\end{figure}

\begin{table}[t]
\centering
\footnotesize
\caption{CSV files in \AgentTorrent, with artifact and repository counts.}
\label{tab:files}
\setlength{\tabcolsep}{3pt}
\begin{tabularx}{\columnwidth}{@{}l>{\raggedright\arraybackslash}Xrr@{}}
\toprule
File & Content & Count & Repos. \\
\midrule
\fn{repos.csv} & metadata and labels & \curFrame & \curFrame \\
\fn{context_files.csv} & e.g., \fn{CLAUDE.md} & \curRowsContext & \curFlagContext \\
\fn{skills.csv} & skill folders & \curRowsSkills & \curFlagSkills \\
\fn{rules.csv} & rule files & \curRowsRules & \curFlagRules \\
\fn{settings.csv} & settings files & \curRowsSettings & \curFlagSettings \\
\fn{subagents.csv} & subagent definitions & \curRowsSubagents & \curFlagSubagents \\
\fn{commands.csv} & custom commands & \curRowsCommands & \curFlagCommands \\
\fn{mcp.csv} & MCP configurations & \curRowsMcp & \curFlagMcp \\
\fn{plugins.csv} & plugin manifests & \curRowsPlugins & \curFlagPlugins \\
\fn{hooks.csv} & hook configurations & \curRowsHooks & \curFlagHooks \\
\fn{lsp.csv} & LSP configurations & \curRowsLsp & \curFlagLsp \\
\midrule
\fn{commits.csv} & credited commits & \curCommits & \curCommitRepos \\
\fn{other_commits.csv} & credited commits & \curOtherCommits & \curOtherRepos \\
\fn{imports.csv} & imports and targets & \curImports & \curImportRepos \\
\fn{planning-documents.csv} & specs, plans, tasks & \curPlanning & \curPlanningRepos \\
\bottomrule
\end{tabularx}
\end{table}

We provide our dataset as CSV files that can be joined by repository name (\cref{tab:files}).
\fn{repos.csv} describes the full sampling frame of \curFrame{} repositories. It contains GitHub metadata and repository IDs, which allow researchers to follow repositories across renames. It also records whether each repository hosts an engineered software project and the reasons for that classification. Classification fields are blank for \curExcluded{} repositories, mainly because their README files were not in English (\cref{sec:sampling}). A \emph{configured repository} is a sampled engineered software project with at least one detected configuration artifact. Of these repositories, \curMultiMechRepos{} (\curMultiMechPct{}) contain at least two configuration mechanisms. Among the \curPlanningRepos{} repositories with planning documents, \curConfigPlanCommitRepos{} (\curConfigPlanCommitPct{}) also contain commits crediting one of the six target tools. For these repositories, \fn{repos.csv} indicates the tools and configuration mechanisms detected. The ten mechanism CSVs record the path, creation date, commit count, and first and last commit hashes of each configuration artifact. Shared configuration artifacts, such as \fn{AGENTS.md}, have \texttt{any} in the \texttt{tool} column because multiple tools can read them. The \curNoSpecific{} repositories with only shared configuration artifacts cannot be attributed to a particular tool from their configuration alone.

For example, researchers can investigate whether tools contribute instructions for the same or other tools. Commits introducing \curContextCredited{} of context files and \curPlanningCredited{} of planning documents credit a tool. In \cref{fig:links}, \emph{Claude Code} receives credit in the commits adding two linked context files and a \emph{Spec Kit} planning document. Researchers can also study instruction dependencies and contradictions: \curImportsDeep{} imports in \curImportsDeepRepos{} repositories connect files beyond the initial instruction file. File dates and commit histories also support trend analyses within a release.

Settings and hooks support studies of permitted commands and checks on agent actions. We also collect \emph{GitHub Agentic Workflows} from \curAgenticRepos{} repositories, exposing tasks delegated through continuous integration. When comparing contributions, researchers can separate accepted review suggestions from other credits: Of \curGeminiCommits{} commits crediting \emph{Gemini}, \curGeminiReviewerOnly{} credit it solely for an accepted review suggestion. For \emph{Copilot}, this applies to \curCopilotReviewerOnly{} of \curCopilotCommits{} credited commits.

We archive our dataset on Zenodo~\cite{baltes2026agenttorrent} with collected and imported files, all heuristics, sampling and collection records, and a README defining every column. The DOI, open formats, GitHub repository IDs, and provenance records support reuse in accordance with the FAIR principles~\cite{wilkinson2016fair}. \AgentTorrent is licensed under CC~BY~4.0 and the pipeline~\cite{baltes2026agenttorrentpipeline} under MIT; collected files and commit messages retain the licenses of their repositories. Our website lets researchers download and browse the dataset and includes a Python usage example: \url{https://se-uhd.de/agenttorrent/}.

\section{Dataset Construction}
\label{sec:construction}

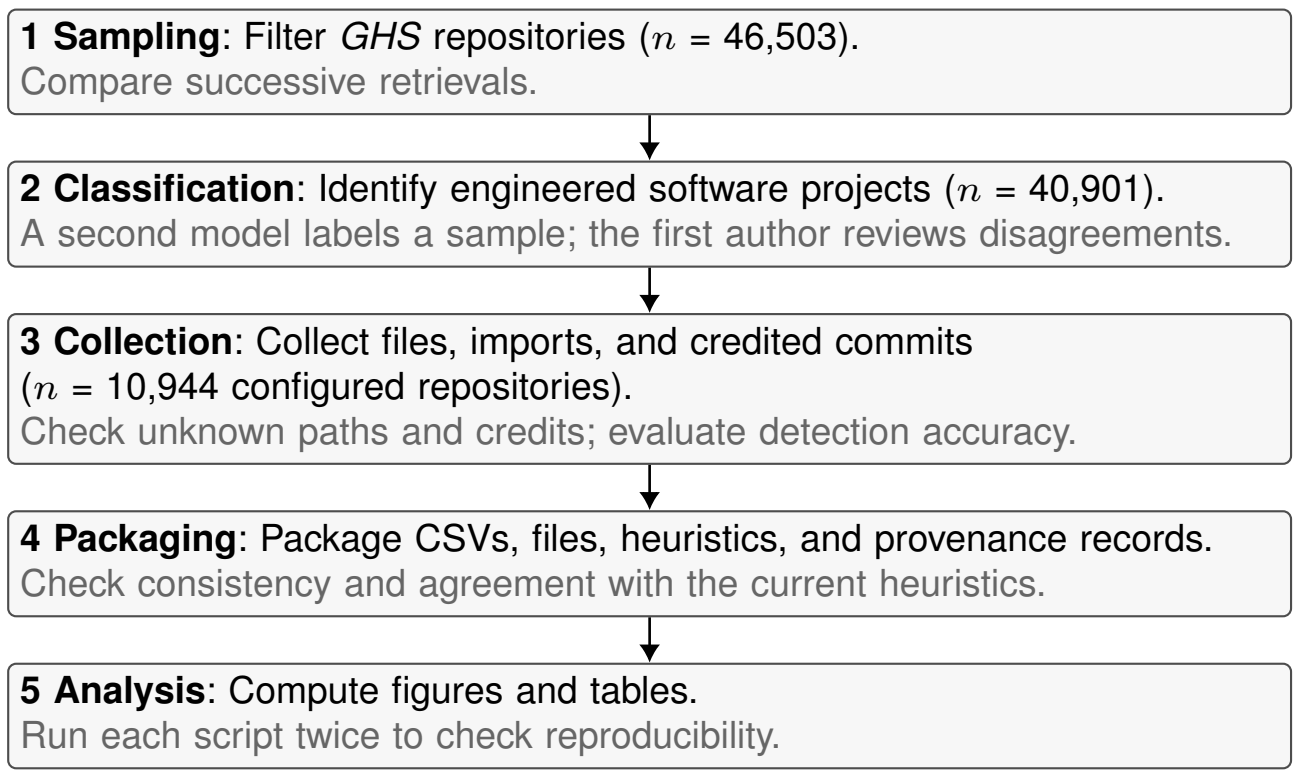
\begin{figure}[t]
\centering
\begin{tikzpicture}[
  font=\sffamily\fontsize{7}{8.5}\selectfont,
  stage/.style={draw=black!70,rounded corners=1.5pt,fill=black!3,align=left,inner sep=0.8mm,text width=0.94\columnwidth},
  flow/.style={-{Latex[length=1.2mm,width=1.4mm]},line width=0.5pt}
]
\node[stage] (a) {\textbf{1 Sampling}: Filter \emph{GHS} repositories \mbox{($n$~=~\curFrame{})}.\\
  {\color{black!60}Compare successive retrievals.}};
\node[stage,below=3mm of a] (b) {\textbf{2 Classification}: Identify engineered software projects \mbox{($n$~=~\curTrue{})}.\\
  {\color{black!60}A second model labels a sample; the first author reviews disagreements.}};
\node[stage,below=3mm of b] (c) {\textbf{3 Collection}: Collect files, imports, and credited commits \mbox{($n$~=~\curConfigured{} configured repositories)}.\\
  {\color{black!60}Check unknown paths and credits; evaluate detection accuracy.}};
\node[stage,below=3mm of c] (d) {\textbf{4 Packaging}: Package CSVs, files, heuristics, and provenance records.\\
  {\color{black!60}Check consistency and agreement with the current heuristics.}};
\node[stage,below=3mm of d] (e) {\textbf{5 Analysis}: Compute figures and tables.\\
  {\color{black!60}Run each script twice to check reproducibility.}};
\draw[flow] (a) -- (b);
\draw[flow] (b) -- (c);
\draw[flow] (c) -- (d);
\draw[flow] (d) -- (e);
\end{tikzpicture}
\caption{The five stages of the \AgentTorrent data collection pipeline. The gray text describes how the output of each stage is checked.}
\label{fig:pipeline}
\end{figure}

Our pipeline automates the five stages listed in \cref{fig:pipeline}. Rerunning it requires GitHub access tokens and an OpenRouter API key, but then operates autonomously.

\subsection{Sampling and Classification}
\label{sec:sampling}

To select active repositories with multiple contributors, we queried \emph{SEART GitHub Search} (\emph{GHS})~\cite{DBLP:conf/msr/DabicAB21} for non-fork repositories with a license, at least \parContributorsMin{} contributors and \parPullsMin{} pull request, created at least \parCreatedMonths{} months before \curEndDate{}, and with a commit in the preceding \parCommittedMonths{} months. \emph{GHS} indexes repositories with at least \ghsMinStars{} stars. The contribution and recency requirements address the risks of personal and inactive projects identified by \citeauthor{DBLP:conf/msr/KalliamvakouGBSGD14}~\cite{DBLP:conf/msr/KalliamvakouGBSGD14}. The pipeline added metadata from GitHub's API and removed unavailable, private, archived, disabled, locked, and forked repositories. Further filters retained OSI-approved software licenses and the \parTopLanguages{} most common primary languages, excluded educational repositories, and required at least the median number of commits and watchers (\parMedianCommits{} commits and \parMedianWatchers{} watchers). We refer to these \curFrame{} repositories as our \emph{sampling frame}.

The classification stage of our pipeline assessed which repositories host an engineered software project~\cite{DBLP:journals/ese/MunaiahKCN17}. We prompted the open-weight model \emph{DeepSeek V4.1 Flash} with each repository's name, description, README, issue and pull request activity, GitHub metadata, and file outline. The model could request summaries of up to \parLinkedPages{} linked pages through the \texttt{read\_linked\_page} tool exposed by an MCP server that we developed. The model evaluated exclusion criteria, such as teaching material or templates, with quoted evidence and checked for at least two engineering practices, such as testing or contributor documentation. The pipeline assigned \texttt{false} when an exclusion criterion applied and \texttt{unclear} when a criterion or the engineering evidence remained uncertain; \texttt{true} required that no exclusion criterion applied and that evidence supported at least two engineering practices. After removing copies with shared head commits, the pipeline retained \curTrue{} repositories as the \emph{sample}, labeled \curFalse{} \texttt{false} and \curUnclear{} \texttt{unclear}, and excluded \curExcluded{} without published labels, mainly because their README files were not in English.

To avoid repeated model requests when updating the dataset, our pipeline reuses a classification when the README, description, model, and classification parameters are unchanged. Keeping the original model responses and request settings lets us reconstruct the published labels without rerunning the model. We archive the prompts, classification checklist, and exclusion criteria with the pipeline on Zenodo~\cite{baltes2026agenttorrentpipeline}.

To evaluate the repository labels assigned by \emph{DeepSeek V4.1 Flash}, \emph{GLM-5.3} independently labeled a random sample drawn from the repositories that the first model had labeled, using the same prompt and tools. We sized the evaluation sample for a target margin of error of \parMargin{} at \parConfidence{} confidence in the estimated agreement rate. We excluded repositories from earlier calibration and label evaluations because we had used those evaluations to refine the prompts and metadata supplied to the models. The models agreed on \evAgreed{} of \evSample{} repositories (\evAgreement{}, \parConfidence{} confidence interval (CI) [\evAgreementLow{}, \evAgreementHigh{}], Cohen's $\kappa$~=~\evKappa{}). The first author reviewed the \evDisagreements{} disagreements and agreed with \emph{DeepSeek V4.1 Flash} on \evRightDeepseek{} repositories, \emph{GLM-5.3} on \evRightGlm{}, and neither model on \evRightNeither{}. This review identified \evDeepseekWrong{} errors in the labels assigned by \emph{DeepSeek V4.1 Flash} within the evaluation sample. Additional errors may remain in that evaluation sample because the first author did not review repositories on which both models agreed. The pipeline repository records the two model labels and the first author's decision for each disagreement. We use feedback from these reviews to continually improve the classification prompts and the metadata supplied to the models.

\subsection{Configuration, Imports, and Agentic Planning Documents}
\label{sec:collection}

A file counts as a configuration artifact when the vendor of an agentic coding tool documents it as repository-level configuration. Our path heuristics encode where and how an agentic coding tool loads configuration. For example, \emph{Cursor} reads \fn{.mdc} files in \fn{.cursor/rules} and ignores \fn{README.md} files in that directory. Each of our \heuPathRows{} path heuristics cites supporting documentation or source code from the vendor and records the date of verification. From \curCollectionFirst{} to \curCollectionLast{}, we cloned the default branches of the repositories in our sample and copied the files matched by our path heuristics to make their contents available for analysis. We followed file renames when reading commit histories to identify the commits that added or last changed each file. Retained commit logs let us apply revised attribution heuristics without cloning repositories again (\cref{sec:currency}). We also recorded paths that resemble configuration but that agentic coding tools do not read, such as \fn{.claude/memory}, with notes explaining why each path does not count as configuration. All \heuAllRows{} heuristics are stored in CSV files, and automated tests require a source for each heuristic.

A file import references another file that the tool loads into its context, such as \fn{@docs/style.md} in a \fn{CLAUDE.md}. Imported files can import further files. Our pipeline follows these chains according to each tool's loading rules, including \emph{Claude Code}'s limit of four successive imports per chain. We compared the imports identified by our parsers with those returned by vendor implementations on \impCasesClaude{} test texts for \emph{Claude Code} and \impCasesGemini{} for \emph{Gemini CLI}. We also collected planning documents, including specifications, plans, and task lists, at paths documented by \heuPlanningFrameworks{} frameworks and tools (e.g., \emph{Spec Kit} and \emph{OpenSpec}), using \heuPlanningRows{} heuristics.

\subsection{Commit Attribution}
\label{sec:attribution}

Commits credit tools through author or committer identities and message text, including co-author trailers. We distinguish tool identities, accepted review suggestions, credits inserted by tools, and user conventions. We used \heuCommitRows{} heuristics to identify commits crediting the six target tools and \heuOtherRows{} to identify commits crediting additional tools. Vendor documentation supports \heuCommitDocRows{} of these heuristics. We derived the remaining heuristics from example commits in GitHub repositories, including credits identified by the checks described in \cref{sec:currency}.

\subsection{Regular Updates and Automatic Checks}
\label{sec:currency}

We plan quarterly releases, starting updates with new \emph{GHS} retrievals and reusing unchanged labels and files.

Our weekly \emph{documentation watch} checks \watchPages{} vendor pages and opens issues in our GitHub repository when those pages change. We review affected heuristics with \emph{Claude Code}. None of the related datasets describes such a documentation check.

While building \AgentTorrent, we reviewed paths in configuration directories and commit credits that none of our heuristics matched. This review revealed additional configuration locations, such as \fn{.claude/workflows}, and undocumented ways of crediting tools. We used these findings to extend our heuristics and will repeat the checks for future releases.

For the collection run published in this release, repository-derived heuristics detected \curSampleOnlySix{} of \curCommits{} credited commits that documentation-based patterns would miss. Before publication, revisions added \revCommitsAdded{} commits, removed \revCommitsRemoved{}, and identified \revNewTools{} additional tools. We reapplied the revised heuristics to that run's retained commit logs without cloning the repositories again. Automated checks verified agreement between tool flags and configuration records, consistent collected commits per repository, and message credits matching the revised heuristics.

\subsection{Evaluation}
\label{sec:validation}

We evaluate detection of configuration artifacts, imports, planning documents, and credited commits. For this release, \emph{Claude Code} agents using \emph{Claude Opus 5.5} labeled samples stratified by tool, mechanism, or heuristic against vendor documentation. Separate agents using the same model relabeled \daSecondPassShare{} of each sample without seeing initial labels, disagreeing on \daPassDisagree{} of \daPassItems{} items. Other agents resolved disagreements and uncertain labels using the explanations and evidence accompanying the detecting heuristics.

Weighting by the number of detections in each stratum, we estimated precision for detected configuration artifacts at \daPathPrec{} (\daConfidence{} CI [\daPathPrecLow{}, \daPathPrecHigh{}], \daPathItems{} sampled detections) and precision for credited commits of the six tools at \daCommitPrec{} (CI [\daCommitPrecLow{}, \daCommitPrecHigh{}], \daCommitItems{} sampled commits). Precision for detected file imports was \daImportPrec{} (CI [\daImportPrecLow{}, 100\%], \daImportItems{} sampled imports); precision for agentic planning documents was \daPlanPrec{} (CI [\daPlanPrecLow{}, 100\%], \daPlanItems{} sampled documents). Our heuristics correctly recorded the type of credit, such as an accepted review suggestion, in \daAttrCorrect{} of \daAttrLabeled{} judgments.

We reviewed repository file lists for configuration missed by our heuristics. In \daConfSampled{} repositories with detections, this review identified \daConfPaths{} configuration files or directories; our heuristics found \daConfPathsFound{} of them. Among \daUnconfSampled{} repositories without detections, \daUnconfMissed{} contained an undetected configuration artifact. For all \daUnconf{} repositories without detections, the extrapolated \daConfidence{} confidence interval extends up to \daMissedHigh{} repositories with missed artifacts. We also sampled \daRecallCommits{} commits uniformly from complete logs: \daRecallCredited{} credited the six tools, and our heuristics found \daRecallFound{}. With only \daRecallCredited{} credited commits, the fraction detected cannot be estimated precisely.

\section{Limitations and Improvements}
\label{sec:limitations}

Detected configuration artifacts in \AgentTorrent do not establish tool use, and AI tools credited in commits do not quantify contributions or capture uncredited usage. Our data collection is limited to default branches of active, licensed, engineered projects in ten languages. Changes in sampling and detection between releases can change total counts even if tool adoption remains unchanged.

Our detection labels are not independent because the same model labeled and resolved samples using heuristics developed with \emph{Claude Code}. Checks of unknown paths can miss shared configuration locations. In future collections, we plan to assess each path heuristic's precision and date when artifacts first appeared at their current paths. Following renames can currently report earlier dates from before a file became configuration.

\section{Conclusion}
\label{sec:conclusion}

\AgentTorrent links configuration artifacts, imported files, planning documents, and credited commits for \curConfigured{} projects, allowing researchers, for example, to identify tools credited for adding instruction files and examine conflicts between planned tasks, instructions, and permissions. We plan quarterly releases and will continuously improve our heuristics, repository classification approach, and tool coverage.

\clearpage
\clubpenalty=0
\widowpenalty=0
\balance
\bibliographystyle{IEEEtranN}
\bibliography{references}

\end{document}